\documentclass[prd,twocolumn,amsmath,amsfonts,amssymb,showpacs,nofootinbib]{revtex4}

\usepackage{epsfig,graphicx,bm,color}
\begin{document}

\title{New Gamma-Ray Contributions to Supersymmetric Dark Matter Annihilation}

\author{Torsten Bringmann}
\email{bringman@sissa.it}

\affiliation{SISSA/ISAS and INFN, via Beirut 2 - 4, I - 34013 Trieste, Italy}
\author{Lars Bergstr\"om}
\email{lbe@physto.se}
\author{Joakim Edsj\"o}
\email{edsjo@physto.se}
\affiliation{Department of Physics, Stockholm University, AlbaNova
 University Center, SE - 106 91 Stockholm, Sweden}

\date{January 25, 2008}


\begin{abstract}
We compute the electromagnetic radiative corrections to all leading
annihilation
processes which may occur in the Galactic dark matter halo, for dark matter in the framework of supersymmetric extensions of the Standard Model (MSSM and mSUGRA), and
present the results of scans over the parameter space that is consistent with present observational bounds on the dark matter density of the Universe.
Although these processes have previously been considered in some special 
cases by various authors, our new general analysis shows novel interesting
results with large corrections that may be of importance, e.g., for searches
at the soon to be launched GLAST gamma-ray space telescope. In particular,
it is pointed out that regions of parameter space where there is a near 
degeneracy between the dark matter neutralino and the tau sleptons, radiative
corrections may boost the gamma-ray yield by up to three or four orders of magnitude, even for neutralino masses considerably below the TeV scale, and will enhance the very characteristic signature of dark matter annihilations, 
namely a sharp step at the mass of the dark matter particle. Since this is 
a particularly interesting region for more constrained mSUGRA models of 
supersymmetry, we use an extensive scan over this parameter space to verify
the significance of our findings. We also re-visit the direct annihilation of neutralinos into photons and point out that, for a considerable part of the parameter space, internal bremsstrahlung is more important for indirect dark matter searches than line signals.
\end{abstract}

\maketitle

\newcommand{\ga}{\gamma}
\newcommand{\be}{\begin{equation}}
\newcommand{\ee}{\end{equation}}
\newcommand{\bea}{\begin{eqnarray}}
\newcommand{\eea}{\end{eqnarray}}
\newcommand{\ds}{{\sf DarkSUSY}}
\newcommand{\py}{{\sf PYTHIA}}
\newcommand{\code}[1]{{\tt #1}}

\hyphenation{}

\section{Introduction}
 During the last few years a strong consensus has emerged about the existence of a sizeable dark matter contribution to the total cosmological energy density. The identification of experimental signatures that eventually may determine the nature of the 
cosmological dark matter is thus becoming ever more important. The 
present estimates \cite{wmap} give the fraction of the critical density of
 cold dark 
matter particles as $\Omega_{CDM}h^2\sim 0.105 \pm 0.013$, where the Hubble 
parameter (scaled in units of 100 km/s Mpc$^{-1}$) is $h\sim 0.70\pm 0.02$.  
Also, 
on the 
scales of galaxies and smaller, a number of methods including measurements 
of rotation
curves as well as gravitational lensing agree well with the predictions from
 N-body calculations of gravitational clustering in cold dark matter cosmologies (see e.g. \cite{vialactea}).

The methods of detection of dark matter (for reviews, see \cite{reviews}) 
can be divided into {\em accelerator} production and detection of missing 
energy (especially at the LHC at CERN, which will start operating some 
time in 
2008), {\em direct  detection} (of dark matter particles impinging on a 
terrestrial detector, with recent impressive upper limits reported by 
\cite{direct}), or {\em indirect 
detection} of particles
generated by the annihilation of dark matter particles in the Galactic halo or in the Sun/Earth.
All these methods are indeed complementary -- it is probable that a signal 
from more than one type of  experiment will be needed to fully identify the 
particle making up the dark matter. The field is just entering very 
interesting times, with the LHC soon
starting and new detectors of liquid noble gases  being developed for 
direct detection. For indirect detection, the satellite PAMELA \cite{pamela}
was launched a year ago and will soon reveal its first sets of data for 
positron and antiproton yields in the cosmic rays \cite{antimatter}. 
 AMANDA \cite{amanda} at the South Pole that has searched for detection of neutrinos
from the centre of the Earth or the Sun \cite{neutrinos}, 
will soon give way to the much larger 
detector IceCUBE \cite{icecube}, and for gamma-rays coming 
from annihilations of dark matter particles in the halo
\cite{gammas}
the space satellite GLAST \cite{glast}, to be launched in 2008, will open up a
 new window to the high-energy
universe, for energies from below a GeV to about 300 GeV. 

One problem with all these discovery methods is that the signal searched for 
may be 
quite weak, with much larger backgrounds in many cases. For indirect detection
through gamma-rays, the situation
may in principle be better, due to (i) the direct propagation from the region of production, without 
significant absorption or scattering; (ii)  
the dependence of the annihilation rate on the square of the dark matter density which 
may
give "hot spots" near density concentrations as those predicted by N-body 
simulations;
(iii) possible characteristic features like gamma-ray lines or steps, 
given by the fact that
no more energy than $m_{\chi}$ per particle can be released in the annihilation of 
two non-relativistic 
dark matter particles (we denote the dark matter particle by $\chi$).

As an example, it was recently shown \cite{idm} that in models of an
 extended Higgs sector,
the line signal from the two-body final states $\gamma\gamma$ and 
$Z\gamma$ could give 
a spectacular signature in the gamma-ray spectrum between 40 and 
80 GeV. On the other hand,
in models of universal extra dimensions (UED) \cite{uedline} or in the theoretically perhaps most favoured, supersymmetric, 
models of dark matter the line
feature is in general not very prominent, except in some particular
 regions of the large parameter space. 
However, it
was early realised that there could be other important spectral 
features \cite{lbe89}, and
recently it has been shown that internal bremsstrahlung (IB) from 
produced charged particles
in the annihilations could yield a detectable "bump" near the 
highest energy for heavy
gauginos or Higgsinos annihilating into $W$ boson pairs, such 
as expected in split supersymmetry
models \cite{heavysusy}.
 In \cite{birkedal}, it was furthermore pointed out that 
IB often can be estimated
by simple, universal formulas and often gives rise to a very prominent step in the spectrum at  
photon energies of $E_\gamma=m_\chi$ (such as in UED models \cite{Bergstrom:2004cy}).

Encouraged by these partial results, we have performed a detailed 
analysis of the importance
of IB in the minimal supersymmetric extension to the standard model (MSSM). We have therefore calculated the IB contributions for all two-particle charged final states from general neutralino annihilations.
Besides 
confirming the mentioned 
partial results for the universal radiative corrections, in particular
those relating to soft and collinear bremsstrahlung, we also point out
interesting cases of model-dependent ``virtual'' brems\-strahlung (i.e. photons emitted from charged virtual particles), see Fig.~1.
We confirm the suspicion expressed already in \cite{lbe89} that
this type of emission may circumvent the chiral suppression, i.e., 
the annihilation
rate being proportional to $m_f^2$ for annihilation into a fermion pair
from an $S$-wave initial state, as is the case in lowest order 
for non-relativistic
dark matter Majorana particles in the Galactic halo (see also \cite{baltz_bergstrom}). Since this enhancement 
mechanism is most prominent in cases where the neutralino is close to 
degenerate with charged sleptons, it is of special importance in the 
so-called stau coannihilation region in models of minimal supergravity (mSUGRA, as implemented in \cite{isajet}). 
We therefore run through an extensive scan over these 
models (based on \cite{baltz_peskin}) and find, indeed, remarkable cases of
enhancement of the gamma-ray rate in the stau coannihilation region, near the maximal possible photon energy
$E_\gamma=m_\chi$.  

Let us stress  that the radiative corrections to the 
main annihilation channels, here computed systematically for the first
time, may turn out to be of utmost importance when fitting gamma-ray data,
e.g. from GLAST, to supersymmetric dark matter templates. Over much of the 
parameter space we have scanned, these corrections give a large factor
of enhancement over the  commonly adopted estimates, especially at the observationally
most interesting, highest energies. More importantly, they add a 
feature, the very sharp step at the dark matter mass, that would distinguish
this signal from all other astrophysical background (or foreground) processes.

\section{Supersymmetric extensions to the standard model}
\label{sec:susy}

Although our calculated electromagnetic radiative corrections will be
applicable to the annihilation of any Majorana dark matter WIMP
(weakly interacting massive particle), we will present results for
the arguably most plausible dark matter candidate; the lightest neutralino
in the MSSM,
which is a linear combination of the superpartners of the gauge and Higgs fields:
\be
  \chi\equiv\tilde\chi^0_1= N_{11}\tilde B+N_{12}\tilde W^3 +N_{13}\tilde H_1^0+N_{14}\tilde H_2^0\,.
\ee
We perform all numerical calculations using the \ds\ code
(see \cite{ds} for our sign conventions and other details).

The parameter $\mu$ is as usual
the Higgsino mass parameter, $\tan\beta$ is the ratio of vacuum expectation
values of the two Higgs doublets, $M_1$, $M_2$ and $M_3$ are the gaugino
mass parameters, $m_A$ is the neutral pseudoscalar Higgs mass.
We are also using parameters $m_0$, $A_t$ and $A_b$, defined through the
simplifying ansatz: ${\bf M}_Q = {\bf M}_U = {\bf M}_D = {\bf M}_E = {\bf M}_L = m_0{\bf 1}$,
${\bf A}_U = {\rm diag}(0,0,A_t)$, ${\bf A}_D = {\rm diag}(0,0,A_b)$, ${\bf A}_E = {\bf 0}$.
Here ${\bf A}$ are soft trilinear couplings
and
${\bf M}$ soft sfermion masses which in general
are $3\times3$ matrices in generation space, but are thus simplified through our ansatz and
encoded in $m_0$, $A_t$ and$A_b$. We do not allow for CP-violating
phases other than the CKM phase of the standard model.
As a natural further simplification, the grand unification condition for the gauge
couplings, leading to $M_1={5\over 3}\tan^2\theta_wM_2\approx {1\over 2}M_2$
is used. For the MSSM scans, we use FeynHiggsFast \cite{feynhiggs} for the Higgs boson masses and decay widths. For each model, we will denote with $m_\chi$ the mass of the lightest neutralino and with $Z_g\equiv\left|N_{11}\right|^2+\left|N_{12}\right|^2$ the gaugino fraction.

As a more restricted, but in some sense more natural, set of parameters we use
those that stem from demanding that the electroweak symmetry be spontaneously
broken by electroweak radiative effects such as appears in minimal supergravity
(mSUGRA) models \cite{msugrarefs}. Also here we use the implementation 
in \ds\ \cite{ds}, which relies on the public code Isajet \cite{isajet} for the solution of the renormalization group equations (RGE) and for the mass spectra.
In these models, parameters are given at the grand unification scale, and are
then (using the RGEs) let to run down to the electroweak scale. 
The range of models with correct
symmetry breaking are usually displayed in the $m_0$-$m_{1/2}$ plane,
where $m_0$ and $m_{1/2}$ are the universal scalar and gaugino masses, respectively,
at the grand unification scale. The additional parameters of mSUGRA models are $\tan \beta$ (as for the MSSM), $A_0$ (which is the common trilinear term at the grand unified scale) and the sign of $\mu$ ($|\mu|$ is determined from the other parameters).

Let us briefly mention those regions in the  $m_0$-$m_{1/2}$ plane that are important from a cosmological point of view,
as they correspond to models with a relic density in accordance with the WMAP value: The \emph{bulk region}
which has low $m_0$ and $m_{1/2}$; the \emph{funnel region}
$m_A\approx 2m_\chi$, where
annihilations in the early universe are enhanced by the presence of the
near-resonant pseudoscalar Higgs boson; the hyperbolic branch or \emph{focus point
region} where $m_0 \gg m_{1/2}$; the \emph{stau coannihilation region}
where $m_{\chi}\approx m_{\tilde\tau}$; and finally the \emph{stop coannihilation region} (arising when $A_0 \ne 0$) where $m_\chi \approx m_{\tilde{t}}$. The stau coannihilation region has
recently been noticed to have favourable properties for indirect detection
rates in antiprotons and gamma-rays \cite{profumo}. In this paper we will show
that, in addition, there is a great enhancement of the high energy gamma-ray
signature in this region.

\section{Internal Bremsstrahlung from WIMP annihilations}
\begin{figure}[t!]
 \includegraphics[width=0.7\columnwidth]{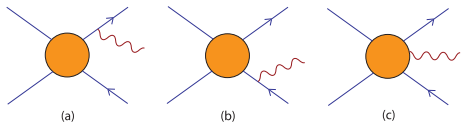}
\caption{\it Types of diagrams that contribute to the first order QED
corrections to WIMP annihilations into a pair of charged particle final states. The leading contributions to diagrams (a) and (b) are universal, referred to as \emph{final state radiation} (FSR),
with a spectral distribution which only depends slightly on the 
final state particle spin and has been calculated, e.g., in  \cite{birkedal}. Internal bremsstrahlung from virtual particles (or \emph{virtual internal bremsstrahlung}, VIB) as in diagram (c), on the other hand, is strongly
dependent on details of the short-distance physics such as helicity properties of the 
initial state and masses of intermediate particles.}
 \label{fig_1}
\end{figure}

\subsection{The general case}

Whenever WIMPs annihilate into pairs of charged particles $X\bar X$, this process will with a finite probability automatically be accompanied by internal 
bremsstrahlung (IB), i.e. the emission of an additional photon in the final state (note that in contrast to ordinary, or external, bremsstrahlung no external electromagnetic field is required for the emission of the photon). As visualized in Fig.~\ref{fig_1}, one may distinguish between photons directly radiated from the external legs (\emph{final state radiation}, FSR) and photons radiated from virtual charged particles (which we will refer to as \emph{virtual internal bremsstrahlung}, VIB). So, to be more specific, \emph{the IB photons will be the total contribution from both FSR and VIB photons}.

If the charged final states are relativistic, FSR diagrams are always dominated by photons emitted \emph{collinearly} with $X$ or $\bar X$.  This is a purely kinematical effect and related to the fact that the propagator of the corresponding outgoing particle,
\be
  D(p)\propto\left((k+p)^2-m_X^2\right)^{-1},
\ee
diverges in this situation. Here, $k$ and $p$ denote the momenta of the photon and the outgoing particle, respectively. The resulting photon spectrum turns out to be of a universal form, almost independent of the underlying particle physics model \cite{Bergstrom:2004cy,birkedal}. Defining the \emph{photon multiplicity} as 
\be
  \frac{dN_{X\bar X}}{dx}\equiv \frac{1}{\sigma_{\chi\chi\rightarrow X\bar X}}\frac{d\sigma_{\chi\chi\rightarrow X\bar X\gamma}}{dx}\,,
\ee
where $x\equiv2E_\gamma/\sqrt{s}=E_\gamma/m_\chi$ and $s$ is the center-of-mass energy, it is given by \cite{birkedal}:
\be
  \label{fsr}
  \frac{dN_{X\bar X}}{dx}\approx\frac{\alpha Q_X^2}{\pi}\mathcal{F}_X(x)\log \left(\frac{s(1-x)}{m_X^2}\right)\,.
\ee
Here, $Q_X$ and $m_X$ are the electric charge and mass of $X$; the splitting function $\mathcal{F}(x)$ depends only on the spin of the final state particles and takes the form
\be
  \mathcal{F}_\mathrm{fermion}(x)=\frac{1+(1-x)^2}{x}
\ee
for fermions and
\be
  \mathcal{F}_\mathrm{boson}(x)=\frac{1-x}{x}
\ee
for bosons. Due to the logarithmic enhancement that becomes apparent in Eq.~(\ref{fsr}), FSR photons are often the main source for IB (note that very near the kinematical endpoint, $x\sim1-m_X^2/s$, it is not sufficient anymore to only keep leading logarithms and one can thus no longer expect Eq.~(\ref{fsr}) to be a good approximation for the actual spectrum).
A prominent example where FSR in this universal form not only dominates IB but in fact the total gamma-ray spectrum from WIMP annihilations, is the case of Kaluza-Klein dark matter \cite{Bergstrom:2004cy}.

In general, one can single out two situations where photons emitted from virtual charged particles may give an even more important contribution to the total IB spectrum than FSR: i) the three-body final state $X\bar X\gamma$ satisfies a symmetry of the initial state that cannot be satisfied by the two-body final state $X\bar X$ or ii) $X$ is a boson and the annihilation into $X\bar X$ is dominated by $t$-channel diagrams. To understand that the first case only leads to an enhancement of VIB, and not of FSR, we recall that the latter is dominated by collinear photons, i.e. the (virtual) final state particles are almost on mass-shell; the two- and three-body final states are thus bound to the same symmetry constraints. The enhancement of the annihilation rate in the second case follows from a closer inspection of the $t$-channel propagator. For non-relativistic WIMPs, it takes the form
\bea
  D_t(p)&\propto&\left((l-p)^2-m_{\tilde X}^2\right)^{-1}\nonumber\\
   &\approx&\left(m_\chi^2-m_{\widetilde X}^2+m_X^2+2m_\chi E_X\right)^{-1}\,,
\eea
where $l$ is the momentum of one of the ingoing WIMPs and $\widetilde X$ denotes the particle that is exchanged in the $t$-channel. If $\chi$ and $\widetilde X$ are almost degenerate in mass, one thus finds an enhancement for small $E_X$ which -- for kinematical reasons -- corresponds to large photon energies $E_\gamma$. Note that this enhancement is less efficient for fermions since the infrared external spinor leg would result in a further suppression factor.
In both cases, and in contrast to the situation for FSR, the resulting spectrum is highly model-dependent. Prominent examples where VIB dominates over FSR as a consequence of these two special situations are neutralino annihilations into leptons \cite{lbe89} and charged gauge bosons \cite{heavysusy}, respectively, to which we will shortly return.
\vspace*{1ex}

\subsection{The neutralino case}

The relevant final states of neutralino annihilations are $W^+W^-$, $W^\pm H^\mp$, $H^+H^-$ and $f\bar f$; analytical expressions for the corresponding rates can be found, e.g., in \cite{reviews}. The inclusion of an additional photon in the final state is straight-forward, though tedious. Here, we are interested in the annihilation of neutralinos today, so we work in the limit of vanishing neutralino velocity -- which greatly simplifies the calculation and the form of the analytical expressions. Our analytical results for arbitrary neutralino compositions agree, in the corresponding limits, with the special situations considered earlier, i.e. pure Higgsino or Wino annihilation into $W^+W^-$ \cite{heavysusy} and photino annihilation into light leptons \cite{lbe89}. Let us stress that, although we state in the following simplified expressions for the photon multiplicities in some special cases, we always use the full expressions in our numerical calculations.

Let us first note that for neutralino annihilations, in contrast to the situation for  Kaluza-Klein dark matter, we cannot in general expect very large contributions from FSR (from external legs). This is because the lightest charged final states, for which the logarithmic enhancement shown in Eq.~(\ref{fsr}) would be most effective, are fermionic -- but the annihilation rate of neutralinos into light fermions is strongly suppressed by a factor $m_f^2/m_\chi^2$ due to the helicity properties of a highly non-relativistic pair of Majorana fermions \cite{Goldberg:1983nd}. On the other hand, as we will demonstrate now,  all the possible final states for neutralino annihilations have the potential of showing considerable VIB contributions. 

For large neutralino masses $m_\chi\gg m_W$ and charginos almost degenerate with the neutralino, e.g., $W^+W^-$ and $W^\pm H^\mp$ final states fall into the second of the categories discussed in the previous subsection.   If the neutralino is an almost pure Higgsino (or Wino), we then find that the photon multiplicity for $W^+W^-$ final states can be well approximated by (see also Ref.~\cite{heavysusy} for a similar result):
\bea
 \frac{\mathrm{d}N_{W^+W^-}} {\mathrm{d}x} &\approx& \frac{\alpha_\mathrm{em}}{\pi}\frac{4(1-x+x^2)^2}{(1-x+\epsilon/2)x} \\
  &&\times\left[\log\left(2\frac{1-x+\epsilon/2}{\epsilon}\right)-1/2+x-x^3\right]\,,\nonumber
\eea
where $\epsilon \equiv m_W/m_\chi$. 
As demonstrated in the next section, the most significant IB contributions from final states with charged gauge bosons to the total annihilation spectrum are, in fact,  often of this form. The full analytical expressions for ${\mathrm{d}N_{W^+W^-}}/{\mathrm{d}x}$ and ${\mathrm{d}N_{W^\pm H^\mp}}/{\mathrm{d}x}$ are rather lengthy and we will therefore not explicitly state them here.  Let us stress again, however, that we of course do use them in our actual calculations.

As mentioned before, the annihilation into light fermions is helicity suppressed; for large photon energies, however, fermion final states containing an additional photon, $f\bar f\gamma$, are not subject to such a suppression. While our full analytical expressions are again rather lengthy, they simplify considerably in the limit of $m_f\rightarrow0$. In this case, and assuming that both sfermions have the same mass, the photon multiplicity is given by

\begin{widetext}
\bea
  \frac{\mathrm{d}N_{f^+f^-}} {\mathrm{d}x} &=& \alpha_\mathrm{em}Q^2_f\frac{\left|\tilde g_R\right|^4+\left|\tilde g_L\right|^4}{64\pi^2} \Big(m_\chi^2 \langle\sigma v\rangle_{\chi\chi\rightarrow f\bar f}\Big)^{-1}\nonumber\\
&&\times(1-x)
\Big\{\frac{4x}{(1+\mu)(1+\mu-2x)}
 -\frac{2x}{(1+\mu-x)^2}
   -\frac{(1+\mu)(1+\mu-2x)}{(1+\mu-x)^3}\log\frac{1+\mu}{1+\mu-2x}\Big\}\,,
\eea
\end{widetext}
where $\mu\equiv m_{{\tilde f}_R}^2/m_\chi^2=m_{{\tilde f}_L}^2/m_\chi^2$ and $\tilde g_RP_L$ ($\tilde g_LP_R$) is the coupling between neutralino, fermion and right-handed (left-handed) sfermion. This confirms the result found in \cite{lbe89} for photino annihilation (while \cite{Flores:1989ru} states a result that  is an overall factor of $2$ larger). Note the large enhancement factor $m_\chi^2/m_f^2$ due to the lifted helicity suppression (from ${\langle\sigma v\rangle}_{\chi\chi\rightarrow f\bar f}\propto m_f^2m\chi^{-4}$), and another large enhancement that appears at high photon energies for sfermions degenerate with the neutralino.

Charged Higgs pairs $H^+H^-$, finally, provide yet another interesting example of the second category discussed in the previous subsection: In the limit $v\rightarrow0$, this final state is not allowed because of  $CP$ conservation. The annihilation into $H^+H^-\gamma$, on the other hand, \emph{is} possible. However, since charged Higgs bosons in most models have considerably larger masses than gauge bosons the enhancement mechanism described in the previous subsection is not as efficient as in the latter case. These final states are thus expected to be of less importance in our context.

In the last part of this section, let us briefly describe how we implemented IB from the various possible final states of neutralino annihilations in \ds. The total gamma-ray spectrum  is given by
\be
  \frac{dN^{\gamma,\mathrm{tot}}}{dx}=\sum_f B_f\left(\frac{dN_f^{\gamma,\mathrm{sec}}}{dx}+\frac{dN_f^{\gamma,\mathrm{IB}}}{dx} + \frac{dN_f^{\gamma,\mathrm{line}}}{dx}\right)\,,
\ee
where $B_f$ denotes the branching ratio into the annihilation channel $f$. The last term in the above equation gives the contribution from the direct annihilation into photons, $\gamma\gamma$ or $Z\gamma$, which result in a sharp line feature \cite{lines}. The first term encodes the contribution from secondary photons, produced in the further decay and fragmentation of the annihilation products, mainly through the decay of neutral pions. 
This ``standard'' part of the total gamma-ray yield from dark matter annihilations shows a feature-less spectrum with a rather soft cutoff at $E_\gamma=m_\chi$.
 In \ds, these contributions are included by using the Monte Carlo code \py\ \cite{PYTHIA} to simulate the decay of a hypothetical particle with mass $2m\chi$ and user-specified branching ratios $B_f$.  In this way, also FSR associated to this decay  is automatically included (the main contribution here comes from photons directly radiated off the external legs, but also photons radiated from other particles in the decay cascade are taken into account). On the other hand, IB from the decay of such a hypothetical particle cannot in general be expected  to show the same characteristics as IB from the actual annihilation of two neutralinos. In particular, and as discussed in length at the beginning of this Section, we expect important VIB contributions in the latter case -- while in the first case there are simply no virtual particles that could radiate photons. We therefore calculate analytically the IB associated to the decay (i.e. FSR from the final legs) and subtract it from $dN_f^{\gamma,\mathrm{sec}}/dx$ as obtained with \py; for $dN_f^{\gamma,\mathrm{IB}}/dx$, we then take the full IB contribution from the actual annihilation process as described before. Hence, this procedure leaves us with corrected \py\ results without FSR on the external legs and our analytical calculation of IB (including FSR and VIB) that we add to this. \footnote{We would like to stress that this prescription is fully consistent since both the original and the corrected IB versions are gauge-invariant separately. Strictly speaking, however, we have only corrected for photons originating directly from the external states and not for those radiated from particles that appear later in the decay cascade. On the other hand, one would of course expect that modifying the energy distribution of the charged particles corresponding to these external legs also affects the further decay cascade. Note, however, that the resulting change in the photon spectrum is a second order effect; more important, for kinematical reasons it does not affect photons at  energies close to $m_\chi$ -- which, as we shall see, are the most relevant. Finally, we observe that our subtraction procedure has only a minor effect on the photon spectrum obtained by \py\ and no practical consequences for the quantities we calculate in the next Section in order to quantify the importance of IB for neutralino annihilations.}

\begin{figure}
\centering
\includegraphics[width=\columnwidth]{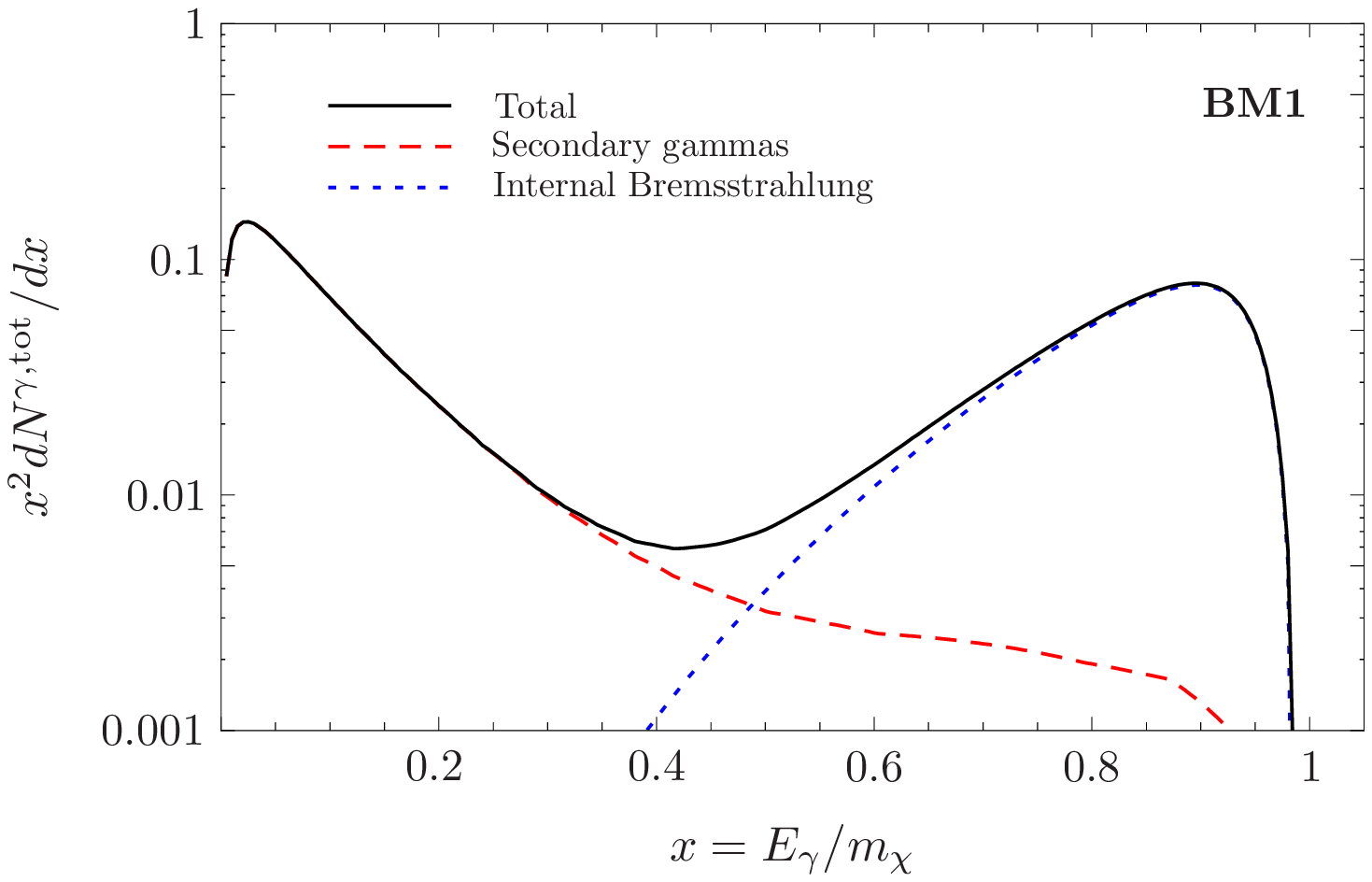}
\includegraphics[width=\columnwidth]{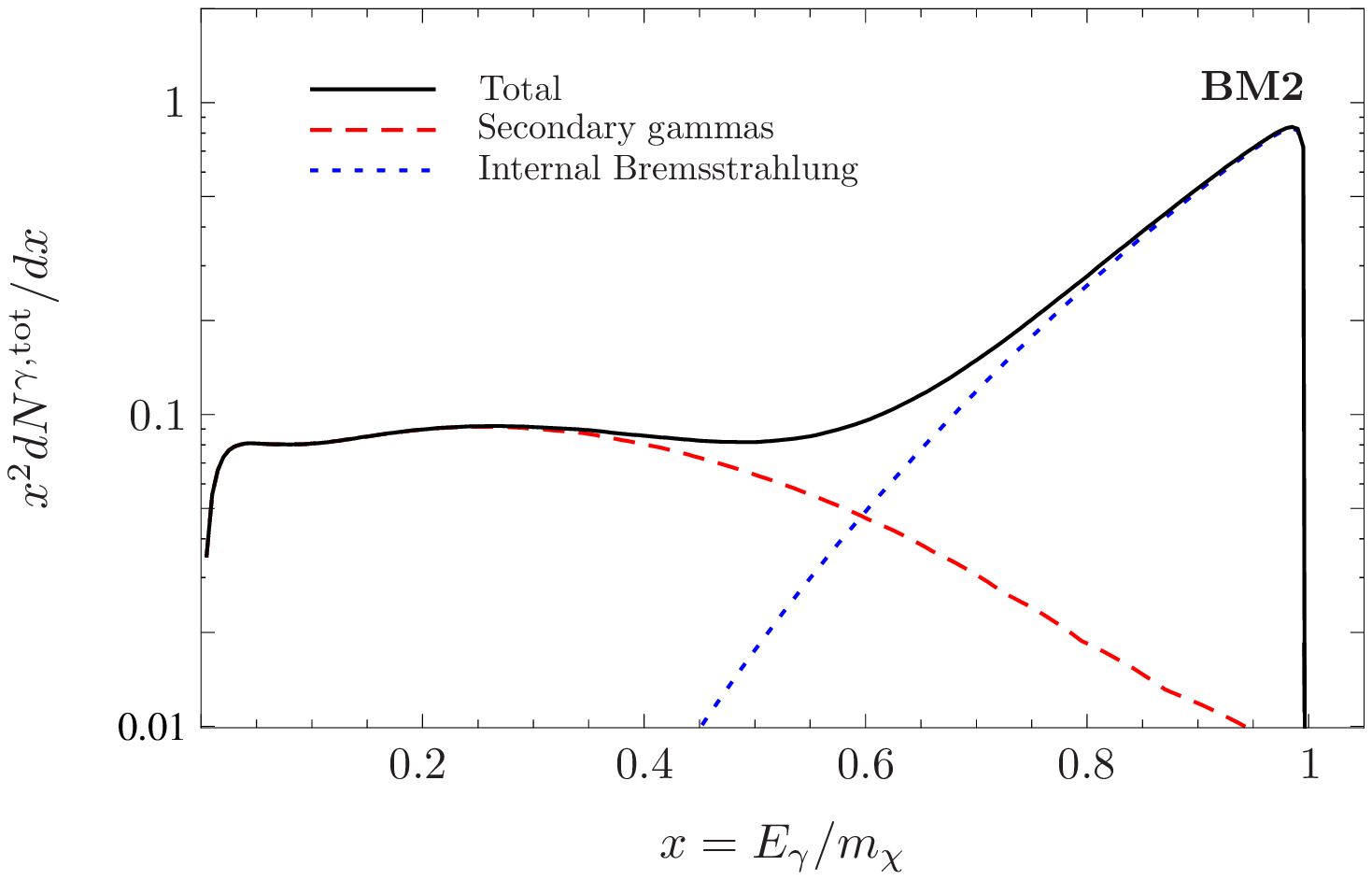}\\
\includegraphics[width=\columnwidth]{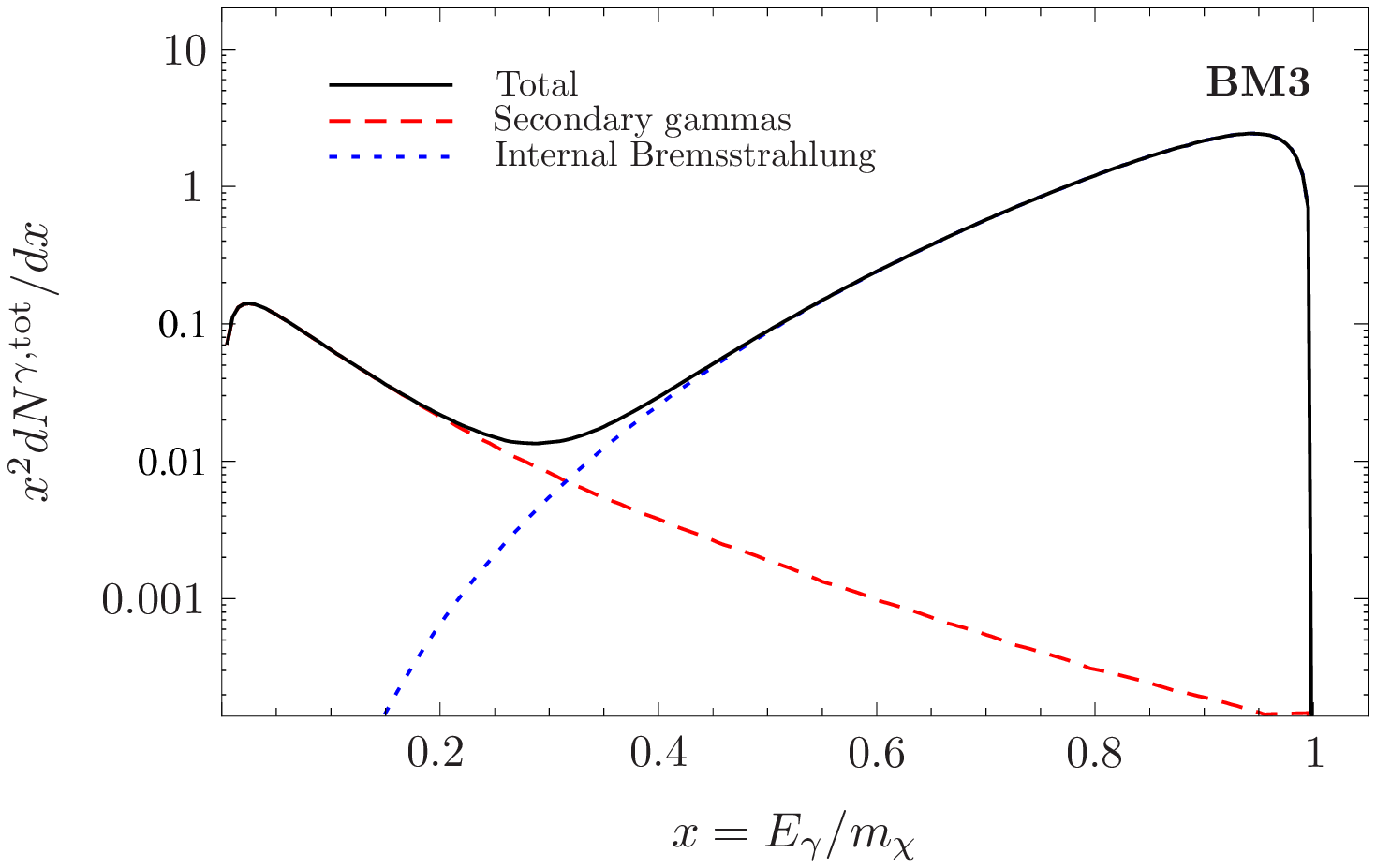}
\includegraphics[width=\columnwidth]{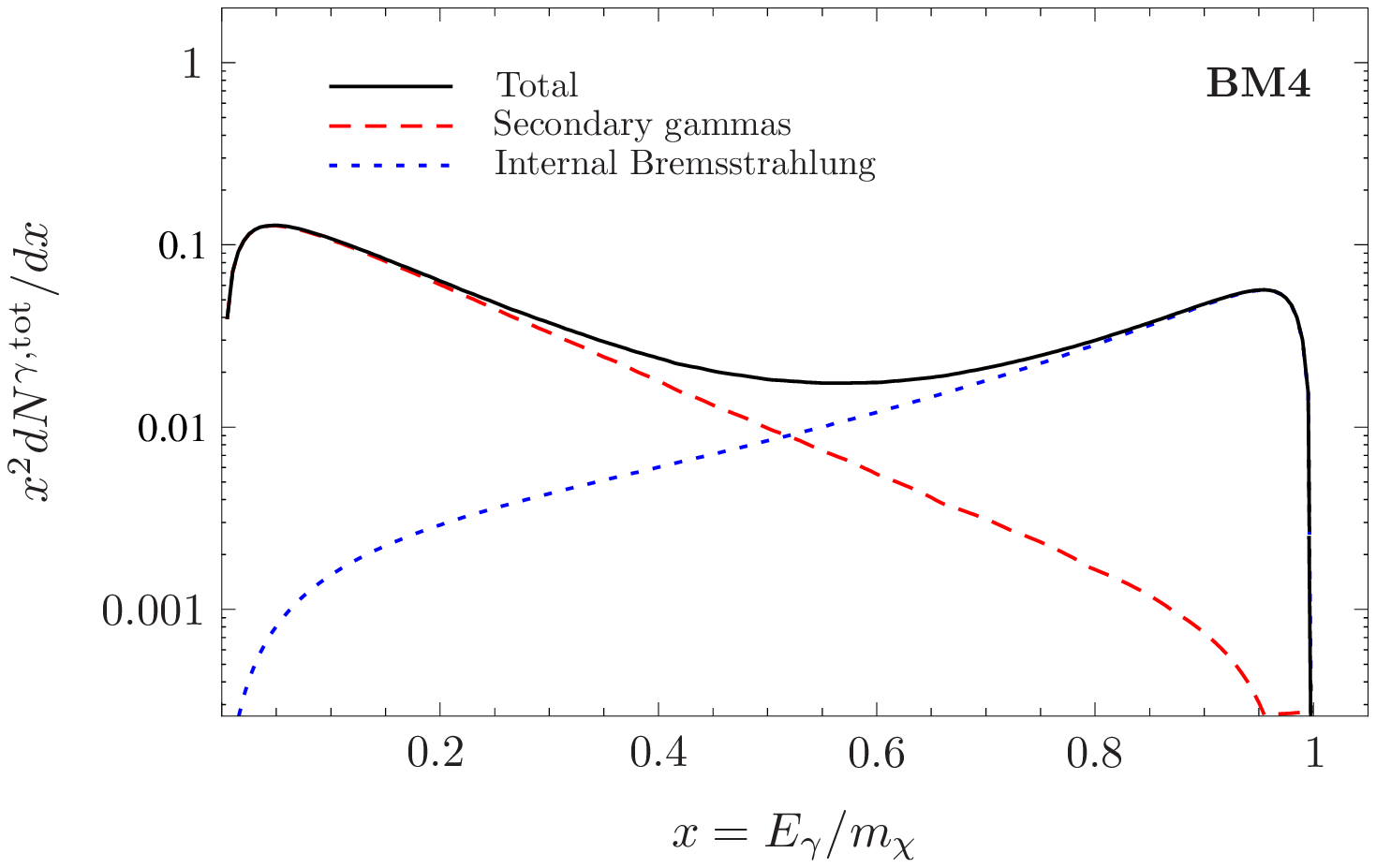}
\caption{From top to bottom, the gamma-ray spectra for the benchmark models defined in Tab.~\ref{benchmark} is shown. The contributions from IB and secondary photons is indicated separately (in these figures, the line signal is not included). }
\label{fig:spec}
\end{figure}

\begin{table*}[t]
    {
    \begin{tabular}{|l||c|c|c|c|c||c|c|c|c|c|c|c|c|}
        \hline
        & $m_0$ & $m_{1/2}$ & $\tan\beta$ & $A_0$ & $\mathrm{sgn}$ & $m_\chi$ & $Z_g/$ & $\Omega h^2$ & $t$-channel & $\cal S$ & IB/ & IB/ \\
    & [GeV] & [GeV] & & [GeV] &$(\mu)$& [GeV] &$(1-Z_g)$& & & & sec.& lines\\
        \hline \hline
       $\!$BM1 & $3700$ & $3060$ & $5.65$ & $\!\!-1.39\!\cdot\!10^4\!$ & $-1$ & $1396$ & $3.0\!\cdot\!10^4$ & $0.082$ & $\tilde t(1406)$ & $8\!\cdot\!10^{-5}\!$ & $19.2$ & $4.5$ \\ 
       $\!$BM2 & $801$ & $1046$ & $30.2$ & $\!\!-3.04\!\cdot\!10^3\!$ & $-1$ & $446.9$ & $1611$ & $0.110$ & $\tilde\tau(447.5)$ & $0.044$ &  $10.6$ & $8.5$ \\ 
       $\!$BM3 & $107.5$ & $\!576.4\!\!$ & $3.90$ & $28.3$ & $+1$ & $233.3$ & $220$ & $0.084$ & $\tilde\tau(238.9)$ & $1.19$ & $\!2.3\!\cdot\!10^3\!$ & $5.0$  \\ 
       $\!$BM4 & $\!2.2\!\cdot\!10^4\!$ & $7792$ & $24.1$ & $17.7$ & $+1$ & $1926$ & $\!\!1.2\!\cdot\!10^{-4}\!\!\!\!$ & $0.11$ & $\tilde \chi^+_1(1996)\!$ & $0.012$ & $10.8$ & $2.1$  \\ 
        \hline
   \end{tabular}}
  \caption{Benchmark models that represent typical regions in the supersymmetric parameter space where IB becomes important. The ``$t$-channel'' entry indicates the main contributing $t$-channel diagram, with the corresponding sparticle and its mass (all masses are given in GeV). ${\cal S}\equiv N_\gamma\frac{\langle\sigma v\rangle}{10^{-29}\mathrm{cm}^3\mathrm{s}^{-1}}\left(\frac{m_\chi}{100\mathrm{GeV}}\right)^{-2}$ is the rescaled flux from IB alone and the last two columns give the ratio of the integrated flux, all above $0.6\,m\chi$, between the new IB contribution and secondary photons as well as the line signals. The main difference between BM2 and BM3 is that in the former neutralinos mainly annihilate into $\tau$ leptons, while in the latter mainly into $t$ quarks. Also, for  BM2 only the $\tau$ final states give an important contribution, while in the second case, even the other leptonic final states contribute considerably (due to near-degenerate slepton masses). For the BM4 model, the IB contribution from the $W^+W^-$ state dominates.
\label{benchmark}}
\end{table*}

 Let us conclude this section by showing in Fig.~\ref{fig:spec} four typical examples of mSUGRA models with particularly pronounced IB features. From top to bottom, they show situations in which IB from $t\bar t$, $\tau^+\tau^-$, all lepton and $W^+W^-$ final states, respectively, dominates the total gamma-ray spectrum from DM annihilations.


\section{Numerical results}

We are now ready to apply the results in the previous section to a set of supersymmetric MSSM and mSUGRA models. For MSSM, we use a set of scans of the 7 parameters mentioned in Sec.~\ref{sec:susy}. These scans are fairly general and partly use the method of Markov Chain Monte Carlo models to focus on models that have the relic density in the WMAP preferred range. In total we have about 200\,000 MSSM models that pass the WMAP relic density constraint and all accelerator constraints (checking experimental
bounds on various masses and branching ratios, such as that of $b\to s\gamma$). For mSUGRA, we use the very large set of scans performed in Ref.\ \cite{baltz_peskin} (also made with a Markov Chain Monte Carlo). This data set contains about 550\,000 models that pass all the constraints.

Since the IB  photons are most distinguished at higher energies (as seen in Fig.~\ref{fig:spec}), we will integrate the flux above $0.6m_\chi$ and compare the flux in this energy range from the flux obtained from secondary photons (arising mainly from $\pi^0$ decays in quark jets). We will also compare it with the monochromatic gamma ray lines \cite{lines} that appear at loop-level.

\begin{figure*}[t!]
\begin{minipage}[t]{0.45\textwidth}
\centering
\includegraphics[width=\textwidth]{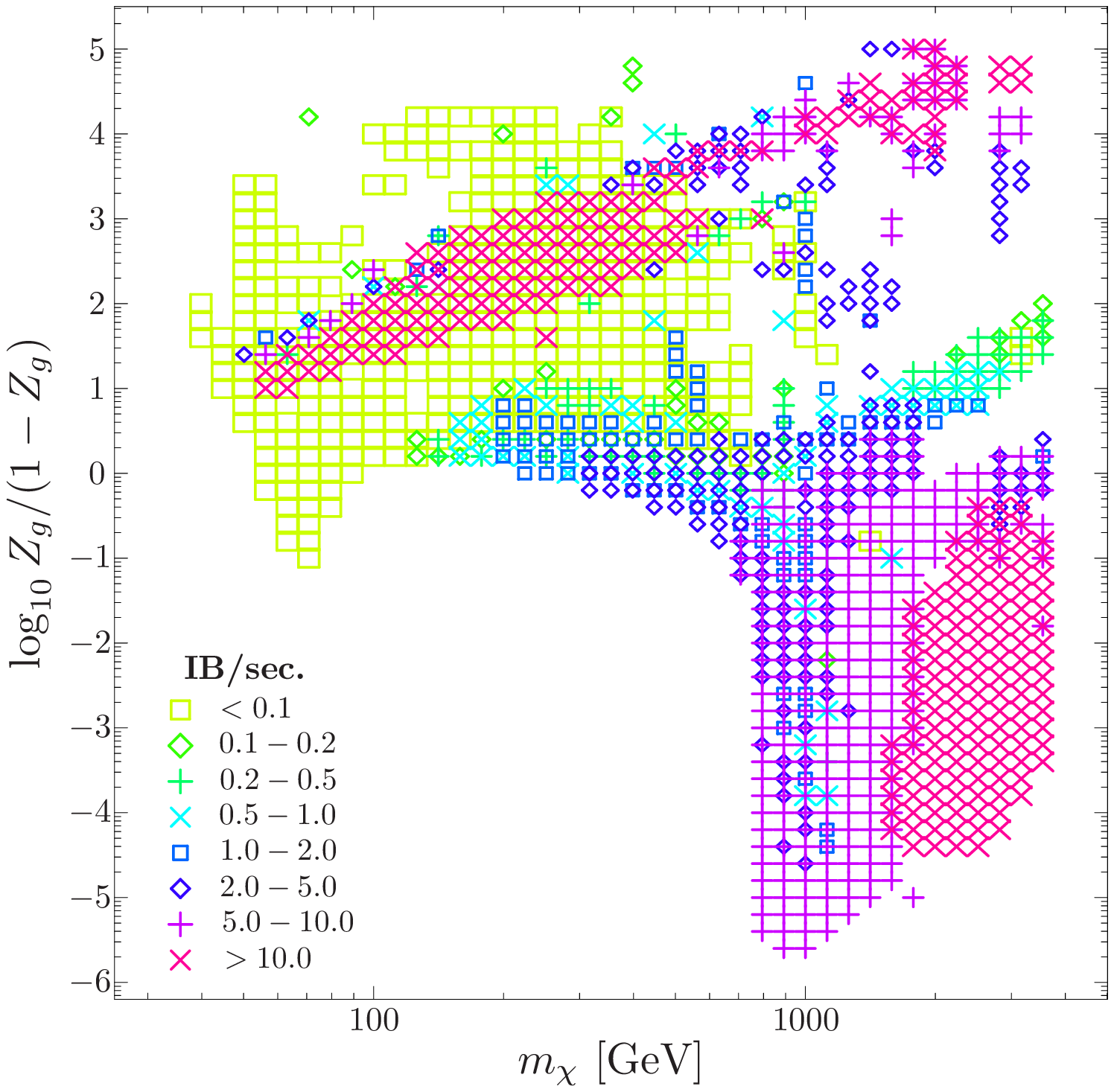}\\
\end{minipage}
\begin{minipage}[t]{0.05\textwidth}
\end{minipage}
\begin{minipage}[t]{0.45\textwidth}
\centering  
\includegraphics[width=\textwidth]{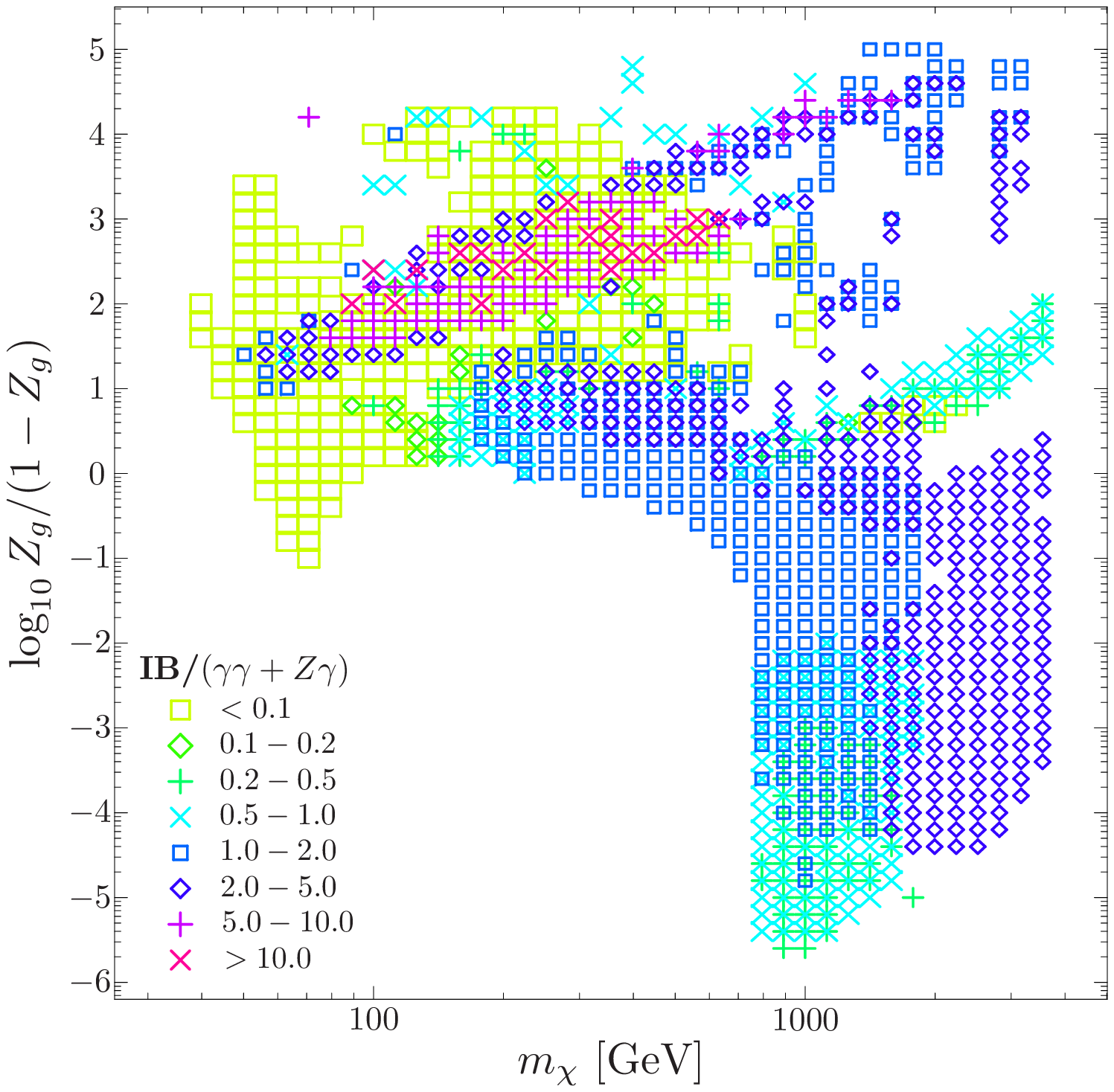}\\
\end{minipage}
\caption{Integrated internal bremsstrahlung flux from supersymmetric dark matter, above $0.6\,m_\chi$, as compared to the ``standard'' continuum flux produced by secondary photons (left) and the flux from both line signals (right).  As for the following figures (\ref{FSRcomp2} and \ref{FSRmSUGRA}), two symbols at the same location always indicate the whole interval between the values corresponding to these symbols.
 Every model considered here features a relic density as determined by WMAP and satisfies all current experimental bounds.
}
\label{FSRcomp1}
\end{figure*}

In Fig.~\ref{FSRcomp1} we show our combined result for both the MSSM and mSUGRA models. In the left panel we show how IB compares with the regular secondary photons from quark jets. We plot this in the $Z_g/(1-Z_g)$ vs $m_\chi$ plane where gauginos (binos and winos) are at the top, Higgsinos are at the bottom and mixed neutralinos are in the middle. In the right panel, we show how  IB photons compare with the monochromatic lines.
It can be clearly seen that in large parts of this plane, IB photons are \emph{the dominating component}, outnumbering both secondary and monochromatic photons. Hence, this effect is very important to include when searching for gamma ray signatures from dark matter.

In Fig.~\ref{FSRcomp1} we looked at the ratio of IB to other gamma ray contributions, let's now turn to the absolute fluxes. We can write the flux observed at earth as
\bea
  \label{eq:flux}
  &&\Phi_\gamma (E,\Delta \Omega,\, \psi) 
    = \\
   &&\qquad
   9.35 \cdot 10^{-14} 
  \, \frac{d{\cal S}}{dE}  \nonumber\times  \langle\,J\left(\psi \right)\,\rangle\,(\Delta\Omega)
  \;\;\rm{cm}^{-2}\;\rm{s}^{-1}\;\rm{sr}^{-1}\,. 
\eea
where we have put all the particle physics into the quantity $\cal S$ and the astrophysics into $J$. $\cal S$ is given by
\be
  \left(\frac{d{\cal S}}{dE}\right)_{\!{\rm IB}\,\gamma}  
  \!\!\!\!=\!\!\!\!\!\!\!  \quad \left( \frac{100\,\rm{GeV}}{m_\chi}\right)^2 \sum_f \left( \frac{v\sigma_f}{10^{-29}\ {\rm cm}^3\; 
   {\rm s}^{-1}}\right)\,\frac{dN_{f}^{\gamma, IB}}{dE}\,,
\ee
where the sum is over all final states $f$.
The astrophysical part (that depends on the chosen halo profile) is given by
\bea
&&\langle\,J\left(\psi\right)\,\rangle\,(\Delta\Omega) =\\
&&\qquad\frac{1} {8.5\, \rm{kpc}}\,\frac{1} {\Delta\Omega}
\int_{\Delta\Omega}\!\!\!d\Omega'\, \mathop{\int_{line}}_{of\,sight}\!\!\!dL\,
\left(\frac{\rho(L,\,\psi')}
{0.3\,{\rm GeV}/{\rm cm}^3}\right)^2\,.\nonumber
\label{eq:jpsi}
\eea
For example, for a standard NFW \cite{nfw} halo profile $\langle J \rangle \simeq 21$ averaged over $\Delta \Omega = 1$ sr towards the galactic centre. This value can be orders of magnitude larger though both for steeper profiles and/or for more concentrated observations towards the galactic centre (i.e.~smaller $\Delta \Omega$ -- as, e.g., for Air Cherenkov Telescopes). The effect of specific halo profiles and a more detailed analysis of the absolute fluxes with respect to the astrophysical background will be left for future work \cite{torstenetal}. We will here focus on the particle physics factor $\cal S$.

\begin{figure*}[t!]
\begin{minipage}[t]{0.32\textwidth}
\centering
\includegraphics[width=\textwidth]{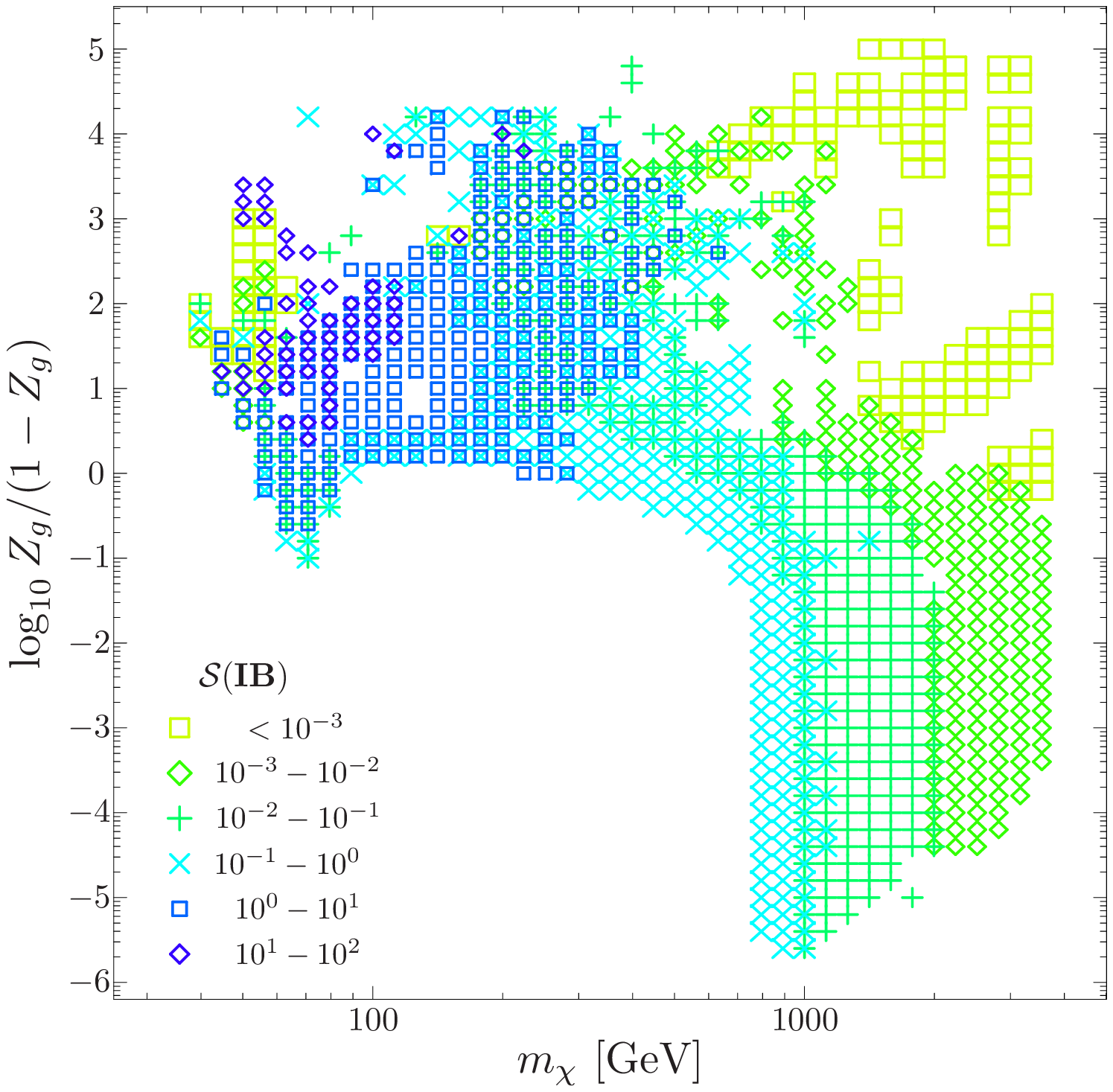}\\
\end{minipage}
\begin{minipage}[t]{0.02\textwidth}
\end{minipage}
\begin{minipage}[t]{0.32\textwidth}
\centering  
\includegraphics[width=\textwidth]{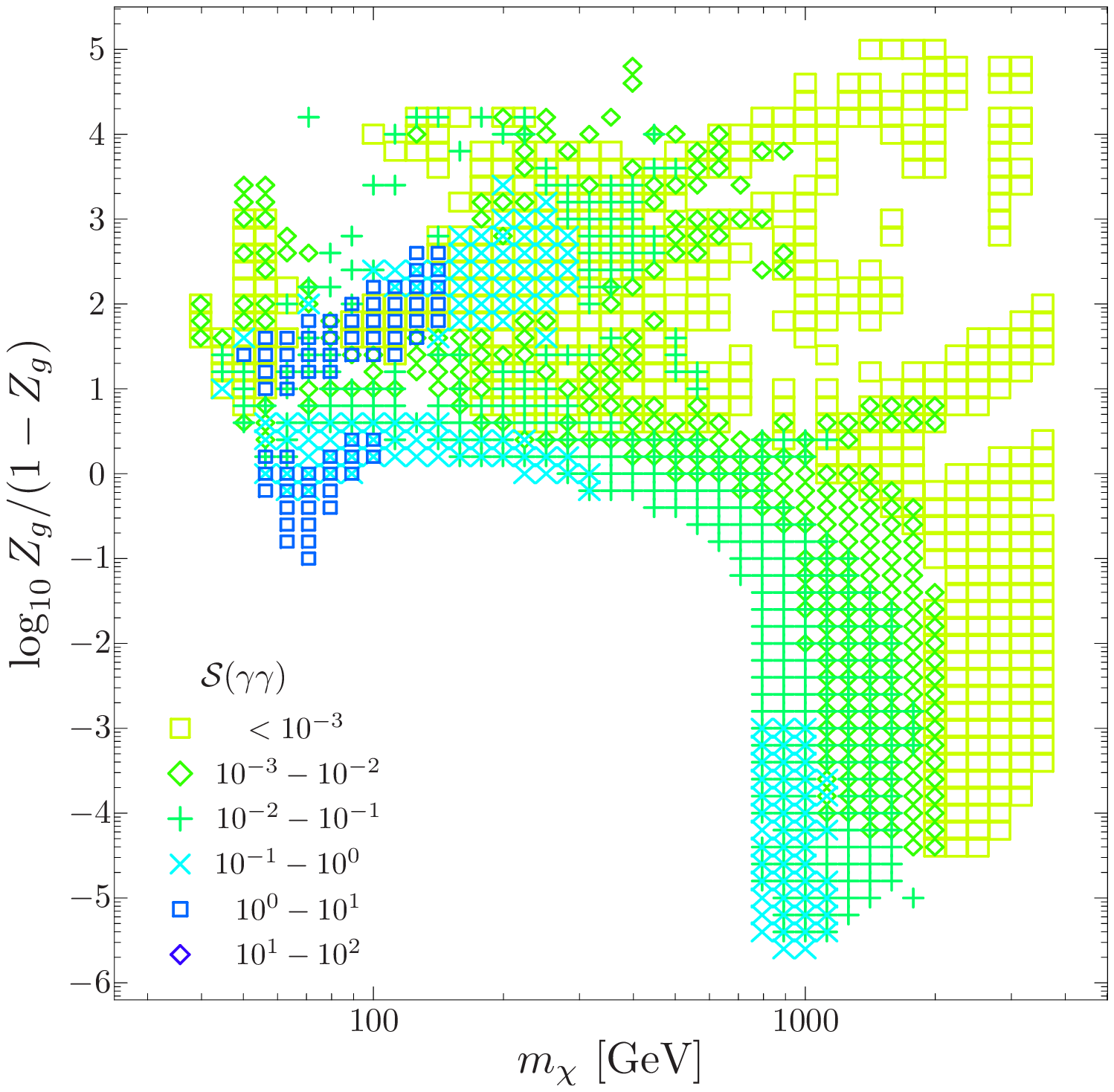}\\
\end{minipage}
\begin{minipage}[t]{0.02\textwidth}
\end{minipage}
\begin{minipage}[t]{0.32\textwidth}
\centering  
\includegraphics[width=\textwidth]{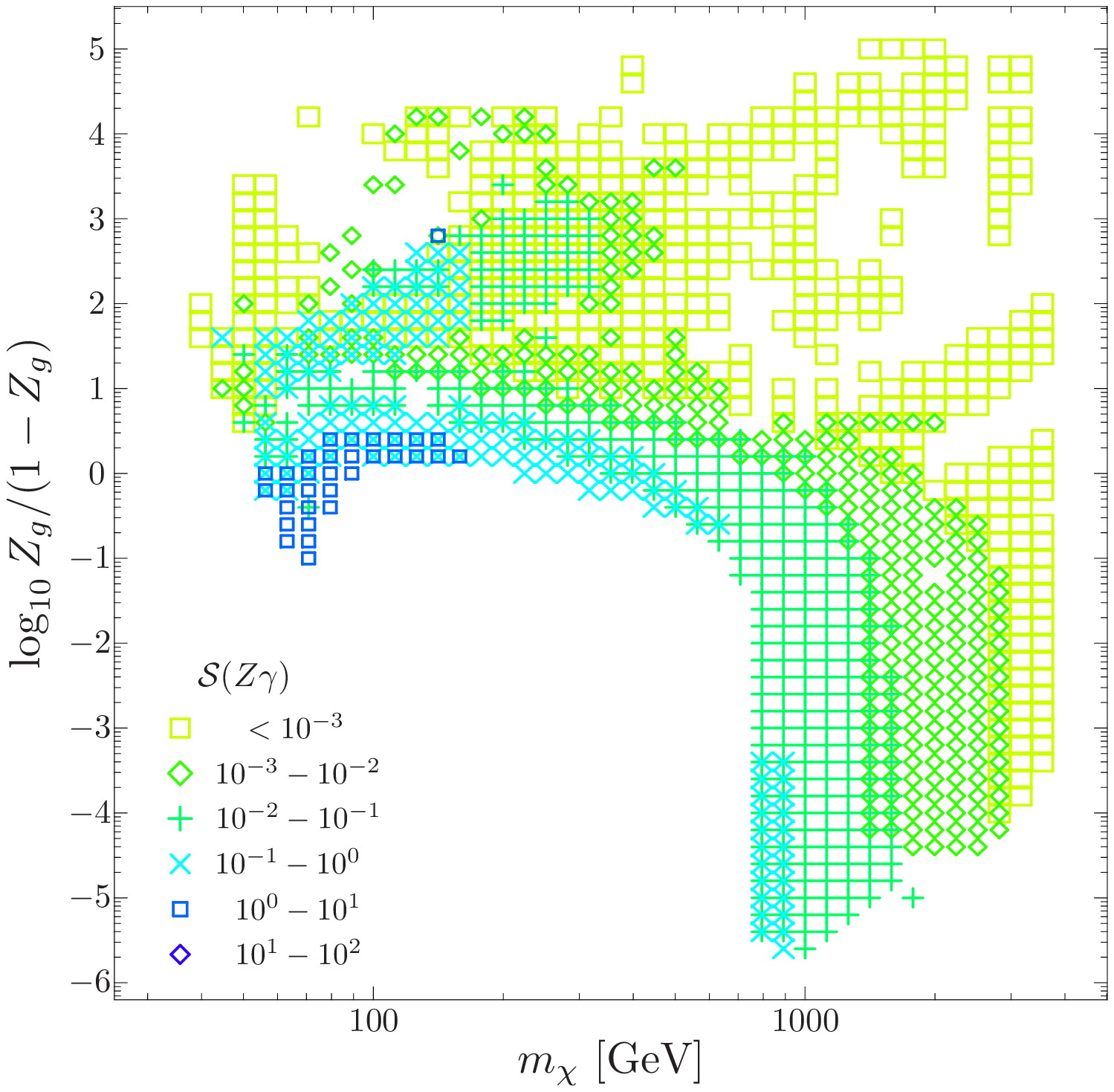}\\
\end{minipage}
\caption{The observationally relevant quantity ${\cal S}\equiv N_\gamma\frac{\langle\sigma v\rangle}{10^{-29}\mathrm{cm}^3\mathrm{s}^{-1}}\left(\frac{m_\chi}{100\mathrm{GeV}}\right)^{-2}$ for IB (left panel) and the line signals (middle and right panel). See text for more details.
}
\label{FSRcomp2}
\end{figure*}

\begin{figure*}[t]
\begin{minipage}[t]{0.32\textwidth}
\centering
\includegraphics[width=\textwidth]{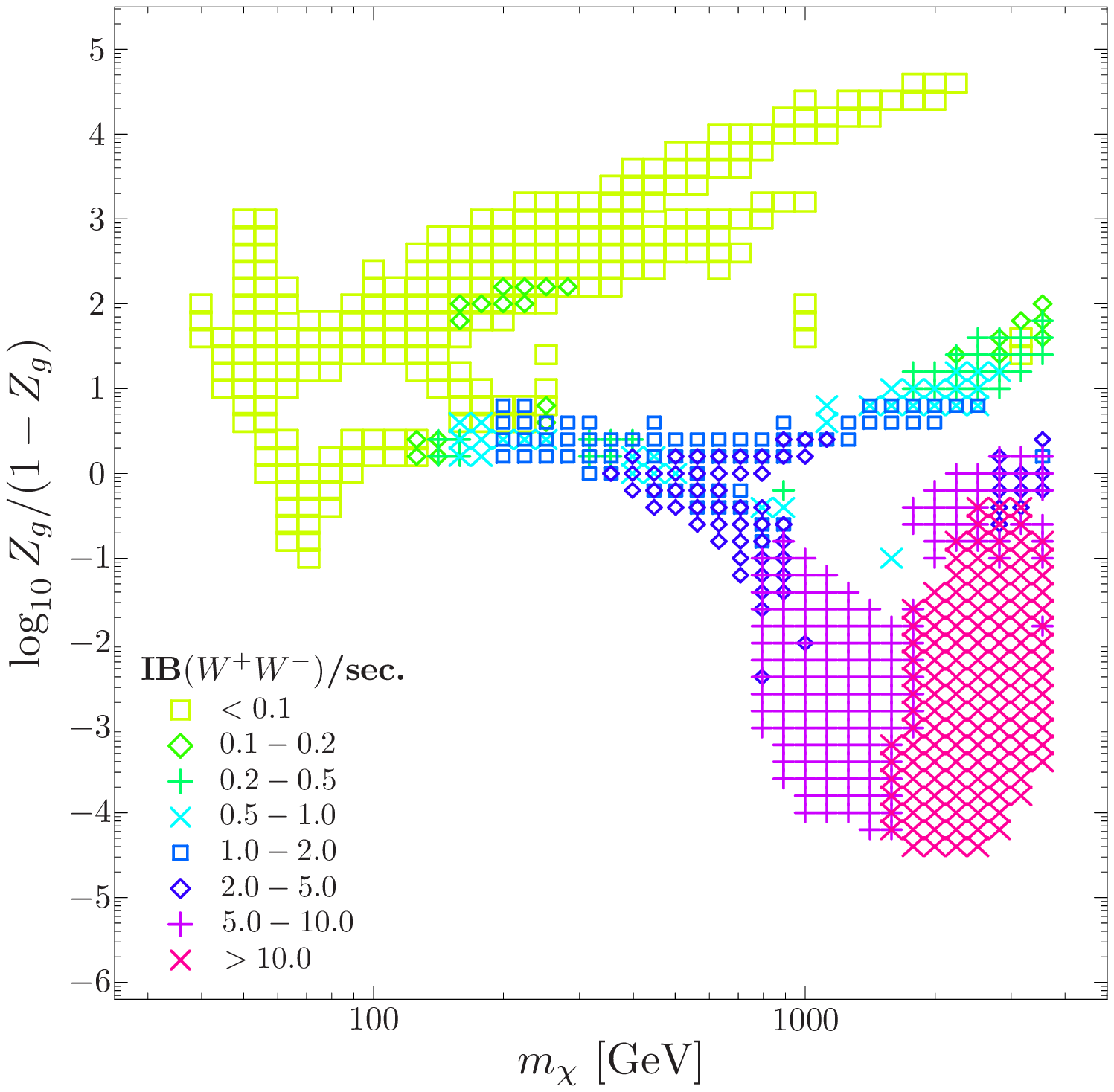}\\
\end{minipage}
\begin{minipage}[t]{0.02\textwidth}
\end{minipage}
\begin{minipage}[t]{0.32\textwidth}
\centering  
\includegraphics[width=\textwidth]{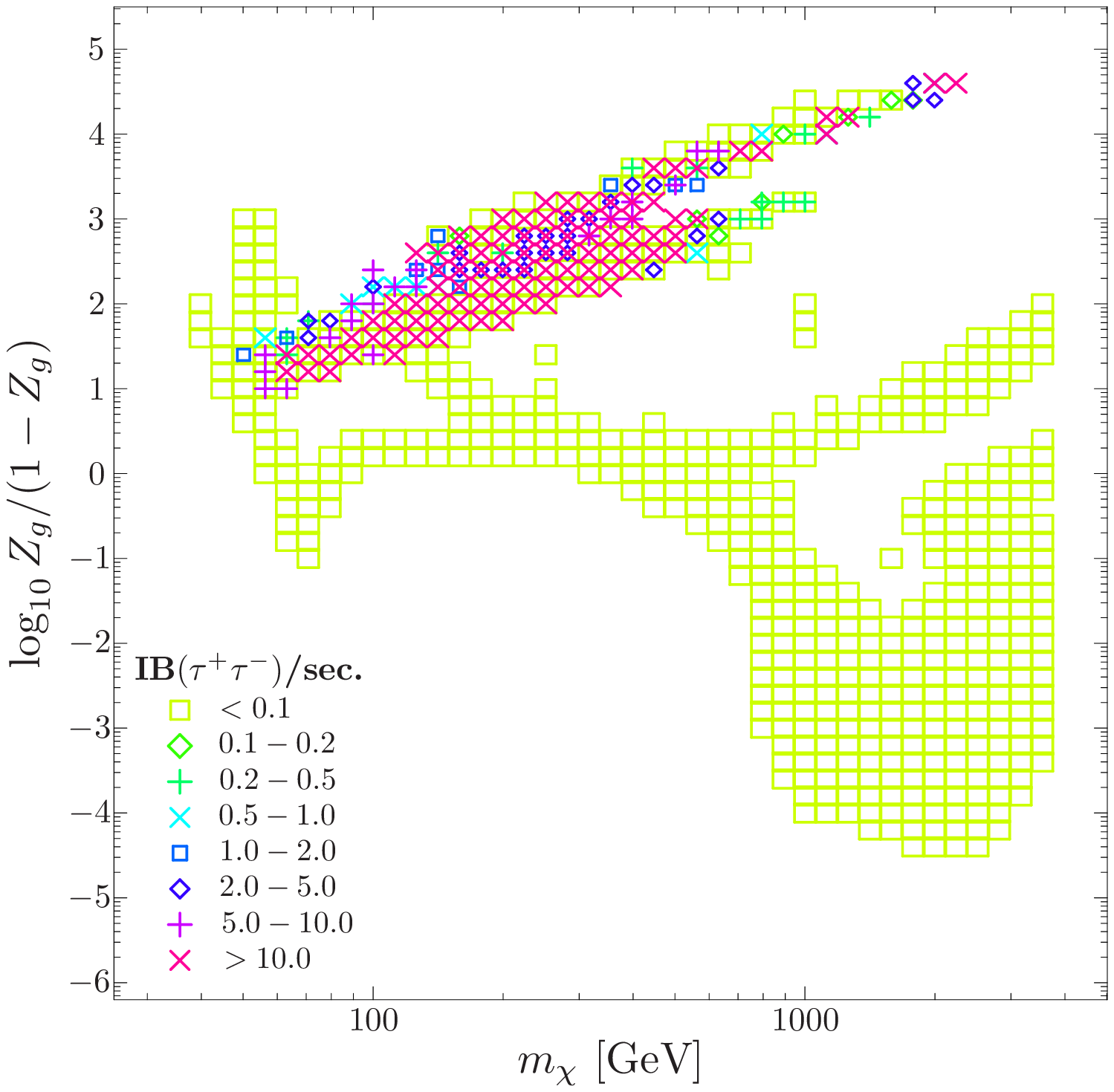}\\
\end{minipage}
\begin{minipage}[t]{0.02\textwidth}
\end{minipage}
\begin{minipage}[t]{0.32\textwidth}
\centering  
\includegraphics[width=\textwidth]{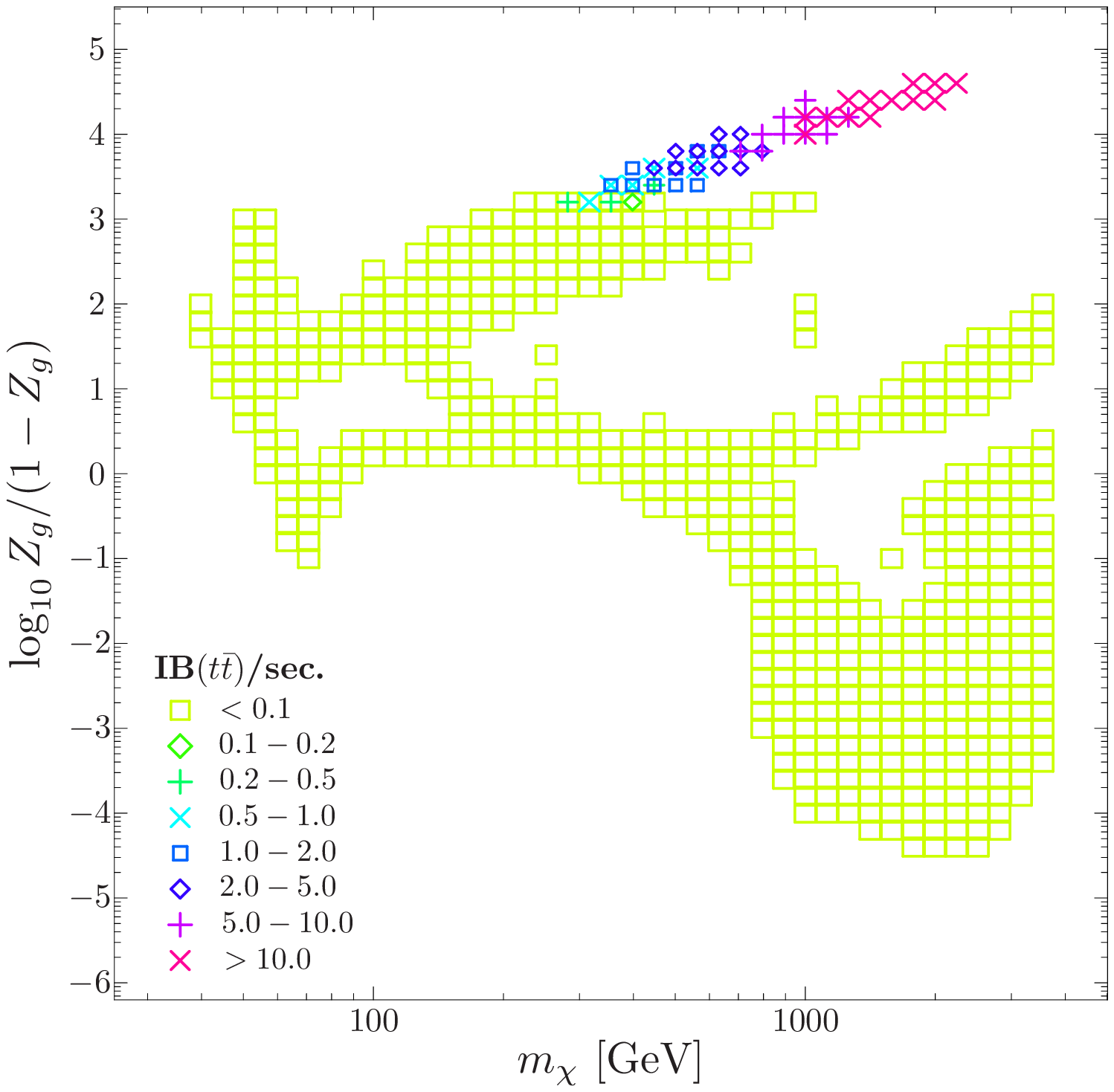}\\
\end{minipage}
\caption{As in the left panel of Fig.~\ref{FSRcomp1}, but now for the individual contributions from various final states of neutralino annihilations in mSUGRA models. IB from light leptons covers a very similar region of the plotted parameter space as that from $\tau$ leptons. All (other) final states not shown here give always IB fluxes less than 10\% of the flux from secondary photons.
}
\label{FSRmSUGRA}
\end{figure*}

In Fig.~\ref{FSRcomp2} we show the quantity $\cal S$, which is $d {\cal S}/dE$ integrated above $0.6\,m_\chi$. In the left panel, we show the yields ${\cal S}$ for the IB contribution, in the middle for monochromatic $\gamma \gamma$ and on the right for $Z \gamma$. In the regions where the IB contribution was the largest in Fig.~\ref{FSRcomp1}, we typically have lower absolute yields. However, there are very pronounced regions, especially at small and intermediate masses, where the IB yields are very high even in absolute terms. We also note that, for neutralino masses in the TeV range, we expect a sizeable increase of the annihilation rate due to non-perturbative effects related to long-distance forces between the annihilating particles \cite{hisano}. These effects have not been taken into account here and would result in a considerable enhancement (by a similar factor) of the quantity $\cal S$ for both line signals \emph{and} IB.

In Fig.~\ref{FSRmSUGRA} we focus on the mSUGRA case and show the contribution relative to the secondary yield of gamma rays for various final states separately. In the left panel, we show the IB yield from the $W^+W^-$ channel, in the middle from the $\tau^+ \tau^-$ channel and in the right from the $t \bar{t}$ channel. Large IB contributions for the $W^+ W^-$ channel occur when a chargino is almost degenerate with the neutralino, as is the case for the focus point region.  Note that due to the grand unification condition, $M_1\approx \frac{1}{2}M_2$, a large gaugino fraction $Z_g$ always means that the neutralino is a \emph{Bino}, with vanishing annihilation rates to $W^+W^-$ or $W^+W^-\gamma$ final states.
The large yields from the $\ell^+ \ell^-$ and $t \bar{t}$ channels, on the other hand, occur when there is a strong degeneracy with the lightest $\tilde{\ell}$ and $\tilde{t}$ respectively. These latter cases occur in the phenomenologically important $\tilde{\tau}$ and $\tilde{t}$ coannihilation regions: in these regimes, coannihilations with  $\tilde{\tau}$ and $\tilde{t}$, respectively are needed to push the relic density down into the WMAP preferred region. Hence, we have a strong mass degeneracy between $\tilde{\chi}$ and $\tilde{\tau}$/$\tilde{t}$ which forces the IB contribution to the gamma yields to be strong. 

As for the other possible final states, we note that the corresponding IB contributions never exceed $10\%$ of the secondary photon flux; these channels are subdominant also for the MSSM models contained in our scan. In fact, from our discussion in the previous section, this is somewhat expected: Charged Higgs bosons, for example, are always heavier than charged gauged bosons, so multi-TeV neutralino masses would be needed for sufficiently large annihilation rates into
$W^\pm H^\mp\gamma$ or $H^+H^-\gamma$ (recalling that the annihilation rate in these cases is enhanced for relativistic final states). IB from light quarks is suppressed by the mass difference between the neutralino and the corresponding squark (as compared to the small mass difference that can be achieved in the stop coannihilation region); down-type quarks, finally, receive a further suppression due to their smaller electric charge.

The main results of our paper may be more easily grasped by looking
at the effect of IB on a small number of 
benchmarks models.
Of course, for the mSUGRA case, it is known that the exact location in 
parameter space of such benchmarks depends very
sensitively on details of the calculation (see e.g.\ \cite{ellis-bench}). 
We therefore define our
own set in Table~\ref{benchmark}, which is very similar to that used by \cite{baltz_peskin} except
that we also include one point in the focus point region (BM4). This set of benchmark models is calculated with ISAJET 7.69 \cite{isajet} together with \ds\ (see \cite{baltz_peskin} for details).
Point BM1 is a model where $A_0$ has been chosen large and negative to make the stop almost degenerate with the neutralino. BM2 is a model where the stau is almost degenerate with the neutralino and in BM3 also the selectron and the smuon are degenerate with the neutralino.
BM4, finally, is in the focus point region, i.e.\ where the lightest chargino is almost degenerate with the lightest neutralino. The main IB characteristics of these benchmark models are summarized in Table~\ref{benchmark}.


\section{Summary and Conclusions}

As can be seen already from our benchmark points in Table~\ref{benchmark}, and in more detail from the
scatter plots in Fig.~\ref{FSRcomp1}, the internal bremsstrahlung effects computed in
this work can be very significant, changing sometimes by more than an order 
of magnitude the lowest-order prediction for the high-energy gamma-ray signal
from neutralino dark matter annihilation. Although some of these enhancements
have been found before \cite{lbe89,heavysusy,birkedal}, this is the first time
the first-order radiative corrections have been computed systematically, for 
all relevant final
states in supersymmetric dark matter models. The resulting enhancements
of the expected fluxes are surprisingly large over significant regions in 
the parameter
space of the MSSM, including the more constrained mSUGRA models. Despite the 
fact that some
large corrections apply to absolute rates that are too small to be of 
practical interest, Fig.~4 shows that the 
quantity ${\cal S}$, which is directly
proportional to the expected signal in gamma-ray detection experiments, also
is significant for the internal bremsstrahlung contribution in large 
regions of parameter space. For 
$m_\chi < 300$ GeV, for example, values of ${\cal S}_{IB}$  greater 
than 0.1 are generic, and
for masses below 100 GeV, values of 1 or higher are 
common, which in very many cases is higher than the corresponding values for
the line signals $\gamma\gamma$ and $Z\gamma$.  One should also bear in mind that the sensitivity of Air Cherenkov Telescopes increases significantly with energy; detectional prospects for a $m_\chi\sim1~$TeV neutralino with ${\cal S}\sim0.01$, e.g., correspond very roughly to those for a $m_\chi\sim100~$GeV neutralino with ${\cal S}\sim0.5$ (see, e.g., \cite{Morselli:2002nw}). In this light, the situation becomes very interesting even for TeV scale Higgsinos, where IB generically contributes more than 10 times as much as secondary photons.

We note that (as anticipated in \cite{lbe89}) helicity suppression and
also CP selection rules of certain final states may be circumvented 
by emitting a photon; this is for example the origin of the very substantial
enhancements of the signal obtained in the stau annihilation
region in mSUGRA models. In this situation, the probability of emitting 
gamma rays vanishes at zero photon energy but increases rapidly at high
energy (see Fig.~\ref{fig:spec}b and \ref{fig:spec}c), which gives a photon ``bump'' at $E_\gamma\approx m_\chi$.  We also note that in less constrained versions of the MSSM than considered here, we expect even more situations where large enhancements of the annihilation signal due to internal bremsstrahlung can be found. An example is heavy Wino dark matter \cite{heavysusy,Chattopadhyay:2006xb}, which becomes possible when relaxing the condition $M_1\approx \frac{1}{2}M_2$ (as realized, e.g., in anomaly mediated supersymmetry breaking scenarios \cite{amsmb}).

Of course, the line signals, in particular $\gamma\gamma$, have the virtue of 
being at the highest possible energy, so in order
to make a more accurate comparison between these and the IB signal computed
here, one would have to model also the expected spectral
shape of possible astrophysical gamma-ray backgrounds and the energy resolution
of the detector.
This is left for future work \cite{torstenetal}. We 
note, however, that in general also the new contributions have a characteristic signature,
({\em cf.} Fig.~2) which can hardly be mimicked by any known astrophysical
gamma-ray source. In fact, in some cases these spectra could even be used by future experiments to distinguish between different dark matter candidates (note that, e.g., the distinction between Kaluza-Klein dark matter and a neutralino in the focus point region like our BM4 point would be possible already with the energy resolution of present Air Cherenkov Telescopes \cite{Bergstrom:2006hk}).

To conclude, we have shown that the commonly neglected first-order radiative
corrections to neutralino dark matter annihilation should definitely be 
taken into account when predicting rates for  gamma-ray telescopes.
In particular, the soon to be launched GLAST space telescope \cite{glast} 
will have an enhanced possibility over what has previously been assumed to 
detect radiation from supersymmetric dark matter annihilation. The routines
needed to compute these new processes will be included in the next release 
of the \ds\ package \cite{ds,joakim}.

\smallskip
\acknowledgments
We wish to thank Piero Ullio, Maxim Perelstein, Michael Gustafsson and Martin Eriksson for useful discussions.
L.B. and J.E. acknowledge support from the Swedish Research Council (VR).
We are grateful to E.A. Baltz for letting us use his extensive data 
base of mSUGRA models.


\end{document}